\documentstyle[12pt]{article}
\vspace{1.8cm} \oddsidemargin 0pt \evensidemargin 0pt  \topmargin
-1.5cm \footheight 26pt \footskip 10mm

\begin{document}
\baselineskip 20pt
\begin{center}
\baselineskip=24pt {\Large\bf The improvement wave
equations of relativistic and non-relativistic quantum mechanics}

\vspace{1cm} \centerline{Xiangyao Wu$^{a}$
\footnote{E-mail:phymath@etang.com}} Wei Han$^{b}$, Gui-Song
Wu$^{a}$, Xiao-Bo Zhang$^{c}$ ,Bingxin Zhang$^{d}$

\vspace{0.8cm}

\noindent{\footnotesize a. \textit{Department of Physics, Huaibei
Coal Industry Teacher's College, Huaibei 235000, China}}\\
{\footnotesize b. \textit{School of Physics and Microelectronics,
Shandong University, Shandong 250100, China}}\\
{\footnotesize c. \textit{School of Materials Science and
Engineering, Shanghai University, Shanghai 200000, China}}\\
{\footnotesize d. \textit{Institute of High Energy Physics,
P.O.Box 918(1), Beijing 100039, China}}
\end{center}

\date{}

\renewcommand{\thesection}{Sec. \Roman{section}} \topmargin 10pt
\renewcommand{\thesubsection}{ \arabic{subsection}} \topmargin 10pt
{\vskip 5mm
\begin {minipage}{140mm}
\centerline {\bf Abstract} \vskip 8pt
\par
\indent In this work, we follow the idea of the De Broglie's
matter waves and the analogous method that Schr\"{o}dinger founded
wave equation, but we apply the more essential Hamilton principle
instead of the minimum action principle of Jacobi which was used
in setting up Schr\"{o}dinger wave equation. Thus, we obtain a
novel non-relativistic wave equation which is different from the
Schr\"{o}dinger equation, and relativistic wave equation including
free and non-free particle. In addition, we get the spin
$\frac{1}{2}$ particle wave equation in potential field.
\end {minipage} }

\vspace*{2cm} {\bf PACS numbers: 03.00.00, 03.65.-w, 03.65.Ta}

\newpage
{\bf 1. Introduction} \vskip 8pt We know De Broglie suggested that
not only does light have a dual nature but material particles also
require a wave-particle description during 1922-23, and he further
noticed correspondences between the classical theory of light and
the classical theory of mechanics. He thought that we may obtain
the wave equation of material particles by comparing the classical
theory of light with the classical theory of material particle
\cite{s1}. Schr\"{o}dinger used De Broglie's idea to obtain the
wave equation of material particles, i.e., Schr\"{o}dinger
equation \cite{s2}. In the following, we also apply De Broglie's
suggestion to research the wave equation of material particles
\cite{s3}. It's well known that Schr\"{o}dinger compared the
Fermat principle with the minimum action principle of Jacobi,
than, from the wave equation of electromagnetic wave obtained the
wave equation of matter particles, i.e., Schr\"{o}dinger equation.
Obviously, the two principles have the similar mathematics form,
but they are different in physics content. So, we begin with the
more essential Hamilton principle, and we find that the principle
can describe not only the geometrical optics but also the
classical mechanics. Applying Hamilton principle instead of the
minimum action principle of Jacobi, we can obtain new wave
equations including time-independent and time-dependent, which are
different from Schr\"{o}dinger equation. With the covariant
Hamilton principle, we can get the relativistic wave equation
including free and external field particle, which extend the
Klein-Gordon equation. Meanwhile, we obtain the relativistic wave
equation of spin $\frac{1}{2}$ particle in potential field.
\\
\\
{\bf 2. Non-relativistic wave equation } \vskip 8pt
 The time-independent wave equation of electromagnetic wave which frequency is $\nu$ is:
\begin{equation}
\nabla^2 \Psi(\vec{r})+\frac{4\pi^2n^2\nu^2}{c^2}\Psi(\vec{r})=0,
\end{equation}
where $n$ is refracting power, $c$ is light velocity and
$\Psi(\vec{r})$ is a component of electromagnetic field $\vec{E}$
and $\vec{B}$. The Eq. (1) describes the wave nature of light such
as interference and diffraction phenomenon of light. When light
transmits at straight line it is described by the geometrical
optics and the geometrical optics is a limiting case of wave
theory of light. Fermat had reduced the laws of geometrical optics
to the principles of 'least-time'. That is, a light ray follows
the path requiring the least time. The Fermat principle is
\begin{equation}
\delta \int n ds=0,
\end{equation}
For a classical material particle, when it moves in potential energy $V(r)$
it can be described by the Hamilton principle
\begin{equation}
\delta \int Ldt=\delta \int\frac{T-V(r)}{v}ds=0,
\end{equation}
where $L=T-V(r)$ is Lagrangian function, $T$, $V(r)$ and
$v=\frac{\sqrt{2m(E-V(r))}}{m}$ are kinetic energy, potential
energy and the velocity of material particle, E is the total
energy. From Hamilton principle, we can deduce Fermat principle,
but we can not get Fermat principle from the minimum action
principle of Jacobi, which is:
\begin{equation}
\delta \int \sqrt{2m(E-V(r))}ds=0,
\end{equation}
In order to obtain the wave equation of material particle we
compare Eq. (2) with (3) not Eq. (2) with (4) which was applied by
Schr\"{o}dinger. Since the Eq. (2) and (3) is concordant we can
think that material particle wave equation is similar to Eq. (1).
In Eq. (1), when $n$ is replaced with
$\frac{(T-V(r))m}{\sqrt{2m(E-V(r))}}$ and the time-independent
wave equation of material particle can be written as follows:
\begin{equation}
\nabla^2
\Psi(\vec{r})+A[\frac{(T-V(r))m}{\sqrt{2m(E-V(r))}}]^{2}\Psi(\vec{r})=0,
\end{equation}
where $A$ is a constant and it can be obtained in the following.
For a free material particle, its potential energy $V(r)=0$ and
total energy $\varepsilon=\frac{p^2}{2m}$, and it is associated
with a plane wave
\begin{equation}
\Psi(\vec{r},t)=\Psi(\vec{r})f(t)=e^{\frac{i}{\hbar}(\vec{p}\cdot\vec{r}-\varepsilon
t)},
\end{equation}
Substitution of Eq. (6) into Eq. (5) gives
\begin{equation}
(\frac{i}{\hbar}\vec{p})^2\Psi(\vec{r})+A\frac{\varepsilon^{2}m^{2}}{2m\varepsilon}\Psi(\vec{r})=0,
\end{equation}
The constant $A$ is
\begin{equation}
A=4/{\hbar}^2,
\end{equation}
From Eq. (5) and Eq. (8)
$$\nabla^2\Psi(\vec{r})+\frac{4}{\hbar^{2}}[\frac{(T-V(r))m}{\sqrt{2m(E-V(r))}}]^{2}\Psi(\vec{r})=0$$
and
\begin{equation}
{-\frac{{\hbar}^2}{2m}\nabla^{2}\Psi(\vec{r})=\frac{(E-2V(r))^{2}}{E-V(r)}\Psi(\vec{r})}.
\end{equation}
as
\begin{equation}
\frac{(E-2V(r))^{2}}{E-V(r)}=\frac{E^{2}-4EV(r)-4V^{2}(r)}{E-V(r)}=E-V(r)-2V(r)+\frac{V^{2}(r)}{E-V(r)}
\end{equation}
Combining these two equations
\begin{equation}
-\frac{{\hbar}^2}{2m}\nabla^{2}\Psi(\vec{r})=(E-V(r)-2V(r)+\frac{V^{2}(r)}{E-V(r)})\Psi(\vec{r})
\end{equation}
The Eq. (11) is the time-independent wave equation of material
particle, and it is different from Schr\"{o}dinger equation
\begin{equation}
-\frac{{\hbar}^2}{2m}\nabla^{2}\Psi(\vec{r})=(E-V(r))\Psi(\vec{r})
\end{equation}
Comparing Eq. (11) with (12), we can find the Eq. (11) has an
additional term $(-2V(r)+\frac{V^{2}(r)}{E-V(r)})$, and this term
can be taken as perturbation in Coulom field, but it is not sure
in the other potential field. Meanwhile, from Eq. (11) we can find
that the momenta $p$ which in the external field can't be made
into operator. But for a free particle $V(r)=0$, the Eq. (11) can
be written as
\begin{equation}
-\frac{{\hbar}^2}{2m}\nabla^{2}\Psi(\vec{r})=E\Psi(\vec{r})
\end{equation}
Obviously, for a free particle, its momenta $p$ can be made into
operator $-i\hbar\nabla$.

The time-dependent wave equation of material
particle can be obtained by time-dependent wave equation of
electromagnetic wave. The time-dependent wave equation of
electromagnetic wave is
\begin{equation}
\nabla^2
\Psi(\vec{r},t)-\frac{n^2}{c^2}\frac{\partial^2\Psi(\vec{r},t)}{\partial
t^{2}}=0,
\end{equation}
According to the same method at the above, we replace $n$ with
$\frac{(T-V(r))m}{\sqrt{2m(E-V(r))}}$, then the time-dependent
wave equation of material particle can be written as
\begin{equation}
\nabla^2
\Psi(\vec{r},t)+A'\frac{(T-V(r))^2m^2}{2m(E-V(r))}\frac{\partial^2\Psi(\vec{r},t)}{\partial
t^{2}}=0,
\end{equation}
For a free material particle, its potential energy $V(r)=0$ and
total energy $\varepsilon=\frac{p^2}{2m}$. Substituting the plane
wave Eq. (6) into Eq. (15), we have
\begin{equation}
(\frac{i}{\hbar}\vec{p})^2\Psi(\vec{r},t)+A'\frac{\varepsilon^2m^2}{2m\varepsilon}
(-\frac{i}{\hbar}\varepsilon)^2\Psi(\vec{r},t)=0,
\end{equation}
The constant $A'$ is
\begin{equation}
A'=-4/\varepsilon^{2},
\end{equation}
From Eq. (15) and Eq. (17)
\begin{equation}
\nabla^2
\Psi(\vec{r},t)-\frac{4}{\varepsilon^{2}}\frac{(T-V(r))^2m^2}{2m(E-V(r))}
\frac{\partial^2\Psi(\vec{r},t)}{\partial t^{2}}=0,
\end{equation}
i.e.
\begin{equation}
\frac{\varepsilon^2}{2m}\nabla^{2}\Psi(\vec{r},t)
=\frac{(E-2V(r))^2}{E-V(r)}\frac{\partial^2\Psi(\vec{r},t)}{\partial
t^{2}},
\end{equation}
The Eq. (19) is time-dependent wave equation of material particle
in external field $V(r)$. Obviously, here the total energy $E$
can't be made into operator.\\
For a free particle, its potential energy $V(r)=0$ and total
energy $\varepsilon=E=\frac{p}{2m}$
\begin{equation}
\frac{\varepsilon}{2m}\nabla^{2}\Psi(\vec{r},t)=\frac{\partial^{2}}{\partial
t^{2}} \Psi(\vec{r},t),
\end{equation}
i.e.
\begin{equation}
\varepsilon(-\frac{\hbar^2\nabla^2}{2m})\Psi(\vec{r},t)=\frac{\partial^{2}}{\partial
t^{2}} \Psi(\vec{r},t),
\end{equation}
According to our conclusion in the front, we know that the momenta
$p$ of a free particle can be made into operator $-i\hbar\nabla$,
so
\begin{equation}
-\frac{\hbar^2\nabla^2}{2m}=\frac{p}{2m}=\varepsilon
\end{equation}
and so
\begin{equation}
\varepsilon^{2}\Phi(\vec{r},t)=-\hbar^{2}\frac{\partial^{2}}{\partial t^{2}}
\Phi(\vec{r},t),
\end{equation}
so that
\begin{equation}
\varepsilon\Phi(\vec{r},t)=i\hbar\frac{\partial}{\partial t}
\Phi(\vec{r},t),
\end{equation}
Obviously, Eq. (24) is the time-dependent wave equation of free
particle. It is accordant with the result of Schr\"{o}dinger and
the energy $\varepsilon$ of free particle can be made into
operator
\begin{equation}
\varepsilon\rightarrow i\hbar\frac{\partial}{\partial t}
\end{equation}
From Eq. (19), we can find that the total energy $E$ of the non-free particle
can't be made into operator.
If the potential energy does not explicitly depend on the time,
the Eq. (19) may be separated by writing
\begin{equation}
\Psi(\vec{r},t)=\Psi(\vec{r})f(t)
\end{equation}
The separated equations are
\begin{eqnarray}
&&\frac{\varepsilon^2}{2m}\frac{E-V(r)}{(E-2V(r))^2}\frac{1}
{\Psi(\vec{r})}\nabla^2\Psi(\vec{r})=C,
\\&& \frac{d^2f(t)}{dt^2}=Cf(t),
\end{eqnarray}
$C$ is a separation constant independent of $\vec{r}$ and $t$.
From Eq. (27), we can get
\begin{eqnarray}
\nabla^2\Psi(\vec{r})=C\frac{2m}{\varepsilon^2}\frac{(E-2V(r))^2}{E-V(r)}\Psi(\vec{r})
\end{eqnarray}
and
\begin{eqnarray}
-\frac{\hbar^{2}}{2m}\nabla^{2}{\Psi(\vec{r})}=C
\frac{(-\hbar^{2})}{\varepsilon^{2}}\frac{(E-2V(r))^{2}}{(E-V(r))}{\Psi(\vec{r})},
\end{eqnarray}
When the coefficient $C$ is
\begin{equation}
C=-\varepsilon^{2}/{\hbar}^2,
\end{equation}
then the Eq. (30) become into the Eq. (9). Substitution of $C$
into Eq. (28) gives
\begin{equation}
\frac{d^2f(t)}{dt^2}=-\frac{\varepsilon^2}{{\hbar}^2}f(t),
\end{equation}
The solution of Eq. (32) is
\begin{equation}
f(t)=B_{1}e^{i\frac{\varepsilon}{\hbar}t}+B_{2}e^{-i\frac{\varepsilon}{\hbar}t},
\end{equation}
The complete solution of Eq. (19) is
\begin{equation}
\Psi(\vec{r},t)=\sum\limits_{n}\Psi_{n}(\vec{r})(B_{1}e^{i\frac{\varepsilon}{\hbar}t}+
B_{2}e^{-i\frac{\varepsilon}{\hbar}t}).
\end{equation}
Where $\Psi_{n}(\vec{r})$ is the eigenfunction of the Eq. (19),
$\varepsilon=\frac{p^2}{2m}$ is the total energy when the
potential energy $V(r)$ is vanished. Now, we can obtain the
conclusions in the following:

(1) The new wave equations of material particle are different from
the Schr\"{o}dinger wave equations.

(2) Comparing Eq. (11) with (12), we can find the time-independent wave equation (Eq. (11))
has an additional term $(-2V(r)+\frac{V^{2}(r)}{E-V(r)})$, and this term can be taken
as perturbation in Coulom field, but it is not sure in the other potential field.

(3) When we compare the time-dependent wave equation (Eq. (19))
with the Schr\"{o}dinger's time-dependent wave equation, we can
find the relation between state and time is
$(C_{1}e^{\frac{i}{\hbar}E_{n}t}+C_{2}e^{-\frac{i}{\hbar}E_{n}t})$
in the Schr\"{o}dinger wave equation, but the relationship is
$(B_{1}e^{\frac{i}{\hbar}\varepsilon
t}+B_{2}e^{-\frac{i}{\hbar}\varepsilon t})$ in our time-dependent
wave equation, i.e., different state according to different
evaluation factor of time in the Schr\"{o}dinger wave equation,
different state according to the same evaluation factor of time in
our wave equation.

(4) We find that the energy $\varepsilon$ and moment $p$ of free
particle can be made into operator $i\hbar\frac{\partial}{\partial
t}$ and $-i\hbar\nabla$. However, the total energy $E$ and momenta
$P$ of the particle in external field can't be made into operator
$i\hbar\frac{\partial}{\partial t}$ and $-i\hbar\nabla$.
\\
\\
{\bf 3. Relativistic wave equation } \vskip 8pt In the following,
we will give the relativistic wave equations of material particle.
Firstly, we should extend the Hamilton principle to covariant
form. The Hamilton principle is \cite{s4}
\begin{equation}
\delta \int L dt=0,
\end{equation}
where $L$ is Lagrange function. The Eq. (35) can be written by
$dt=\gamma d\tau$ and $ds=c d\tau$ as
\begin{equation}
\delta \int \frac{\gamma}{c}L ds=0,
\end{equation}
where $\gamma=\frac{1}{\sqrt{1-\frac{u^2}{c^2}}}$, $u$ is particle
velocity, $c$ is light velocity, $d\tau$ is static time and $ds$
is four-dimension differential interval. Obviously, the Eq. (36)
is covariant when $\gamma L$ is Lorentz-invariant.
\\
For a free particle \cite{s5}
\begin{equation}
\gamma L=-m_0c^2
\end{equation}
and for photon, we suppose
\begin{equation}
\gamma L=-E_{s}=constant
\end{equation}
where $E_{s}$ is the static energy of a photon. From Eq. (36) and Eq. (38)
\begin{equation}
\delta \int ds=0,
\end{equation}
The Eq. (39) is covariant form of the Fermat principle $(n=1)$,
and none but $n=1$, the time-dependent wave equation of
electromagnetic wave is covariant.
\\
From the Eq. (1), when $n(n=1)$ is replaced with $-m_0c^2$, we can
obtain relativistic wave equation of free particle
\begin{equation}
\nabla^2 \Psi(\vec{r})+B(-m_{0}c^{2})^{2}\Psi(\vec{r})=0,
\end{equation}
For a free particle
\begin{equation}
\Psi(\vec{r},t)=\Psi(\vec{r})f(t)=e^{\frac{i}{\hbar}(\vec{p}\cdot\vec{r}-Et)},
\end{equation}
where $E$ is the energy of free particle.
From Eq. (40) and Eq. (41)
\begin{equation}
(\frac{i}{\hbar}\vec{p})^2+B m_0^2c^4=0,
\end{equation}
The constant $B$ is
\begin{equation}
B=\frac{p^{2}}{\hbar^{2}m_{0}^{2}c^{4}}=\frac{E^{2}-E_{0}^{2}}
{E_{0}^{2}\hbar^{2}c^{2}}
\end{equation}
Substitution of Eq. (43) into Eq. (40) gives
\begin{equation}
\nabla^2 \Psi(\vec{r})+\frac{E^{2}-E_{0}^{2}}{E_{0}^{2}
\hbar^{2}c^{2}}E_{0}^{2}\Psi(\vec{r})=0,
\end{equation}
i.e.
\begin{equation}
(E_{0}^{2}-\hbar^{2}c^{2}\nabla^2)\Psi(\vec{r})=E^{2}\Psi(\vec{r}),
\end{equation}
Comparing with
\begin{equation}
(E_{0}^{2}+c^{2}p^2)\Psi(\vec{r})=E^{2}\Psi(\vec{r}),
\end{equation}
Making momentum $p$ into operator $-i\hbar\nabla$, we can obtain
the Eq. (45) by Eq. (46).

In the following, we give the time-dependent relativistic wave
equation of free particle by replacing $n(n=1)$ with $-m_0c^2$ in
Eq. (13), then
\begin{equation}
\nabla^2
\Psi(\vec{r},t)+B^{'}(-m_{0}c^{2})^2\frac{\partial^2}{\partial
t^2}\Psi(\vec{r},t)=0,
\end{equation}
Similarly, substitution of the Eq. (40) into Eq. (47) gives
\begin{equation}
B^{'}=-\frac{p^{2}}{E_{0}^{2}E^{2}}=-\frac{p^{2}}{E_{0}^{2}(E_{0}^{2}+c^{2}p^2)}
\end{equation}
from Eq. (48) and Eq. (47)
\begin{equation}
\nabla^2 \Psi(\vec{r},t)-\frac{p^{2}}{(E_{0}^{2}+c^{2}p^2)}
\frac{\partial^2}{\partial t^2}\Psi(\vec{r},t)=0,
\end{equation}
Making $p$ into operator $-i\hbar\nabla$, we can obtain the
quantum wave equation as following:
\begin{equation}
-\hbar^2\nabla^2 \frac{\partial^2}{\partial t^2}\Psi(\vec{r},t)=
(-c^{2}\hbar^{2}\nabla^{2}+m_{0}^{2}c^{4})\nabla^2 \Psi(\vec{r},t),
\end{equation}
define the wave function $\Phi(\vec{r},t)$
\begin{equation}
\Phi(\vec{r},t)=\nabla^{2}\Psi(\vec{r},t),
\end{equation}
as $\nabla^{2}$ and $\frac{\partial^{2}}{\partial t^{2}}$ are compatible
and so
\begin{equation}
-\hbar^2\frac{\partial^2}{\partial t^2}\Phi(\vec{r},t)=
(-c^2\hbar^2\nabla^{2}+m_{0}^{2}c^{4})\Phi(\vec{r},t),
\end{equation}
Obviously, the Eq. (52) which we obtain is the same as Klein-Gordon
equation.

In the following, we consider a particle in the external potential
field $V(x_{\mu})$. In order to get time-independent relativistic
wave equation, $\gamma L$ should be Lorentz-invariant, and it can
be constructed
\begin{equation}
\gamma L =-m_{0}c^2-V(x_{\mu}),
\end{equation}
where $V(x_{\mu})$ is the Lorentz-invariant potential. Replacing
$n(n=1)$ with $\gamma L$ in Eq. (1), we can get
\begin{equation}
\nabla^2\Psi(\vec{r})+D (-m_{0}c^{2}-V(x_{\mu}))^2
\Psi(\vec{r})=0,
\end{equation}
For a free particle
$$V(x_{\mu})=0,$$
\begin{equation}
\Psi(\vec{r},t)=\Psi(\vec{r})f(t)=e^{\frac{i}{\hbar}({\vec{p}
\cdot\vec{r}}-\varepsilon t)},
\end{equation}
Substitution of Eq. (55) into Eq. (54) gives
\begin{equation}
(\frac{i}{\hbar}\vec{p})^{2}+Dm_{0}^{2}c^{4}=0,
\end{equation}
then
\begin{equation}
D=\frac{p^{2}}{\hbar^{2}m_{0}^{2}c^{4}}=\frac{\varepsilon^2-E_{0}^2}
{\hbar^2c^2E_{0}^2},
\end{equation}
where $\varepsilon=E-V(x_{\mu})=\sqrt{c^2p^2+E_0^2}$, and $E$ is
the total energy of particle in the potential field.

From Eq. (54) and Eq. (57)
\begin{equation}
\nabla^2\Psi(\vec{r})+\frac{(\varepsilon^2-E_{0}^2)}{\hbar^2c^2E_{0}^2}
(E_{0}+V(x_{\mu}))^2\Psi(\vec{r})=0,
\end{equation}
and
\begin{equation}
-c^2\hbar^2\nabla^2\Psi(\vec{r})=[(E-V(x_{\mu}))^2-E_{0}^2][1+\frac{V(x_{\mu})}{E_{0}}]^2\Psi(\vec{r}),
\end{equation}
the Eq. (59) is the time-independent wave equation of the material particle in the external
potential field.

In the following, we can get time-dependent relativistic wave equation of the material
particle in the external potential field by replacing $n(n=1)$ with $\gamma L$ in Eq. (13).
\begin{equation}
\nabla^2\Psi(\vec{r},t)+D^{'}
(-m_{0}c^{2}-V(x_{\mu}))^2 \frac{\partial^2}{\partial t^2}\Psi(\vec{r},t)=0,
\end{equation}
from Eq. (55) and Eq. (60)
\begin{equation}
(\frac{i}{\hbar}\vec{p})^{2}+D^{'}m_{0}^{2}c^{4}(-\frac{i}{\hbar}\varepsilon)^2=0,
\end{equation}
then
\begin{equation}
D^{'}=-\frac{p^{2}}{E_{0}^2\varepsilon^{2}}=-\frac{p^2}{E_{0}^2(E_{0}^2+c^2p^2)},
\end{equation}
Substitution of Eq. (62) into Eq. (60) gives
\begin{equation}
\nabla^2\Psi(\vec{r},t)-\frac{(E_{0}+V(x_{\mu}))^2}{E_{0}^2(E_{0}^2+c^2p^2)}p^2
\frac{\partial^2}{\partial t^2}\Psi(\vec{r},t)=0,
\end{equation}
Making momentum $p$ into operator $-i\hbar\nabla$, we can obtain
\begin{equation}
(E_{0}^2-c^2\hbar^2\nabla^2)\nabla^2\Psi(\vec{r},t)+\frac{(E_{0}+V(x_{\mu}))^2}{E_{0}^2}\hbar^2
\nabla^2\frac{\partial^2}{\partial t^2}\Psi(\vec{r},t)=0,
\end{equation}
define the wave function $\Phi(\vec{r},t)$
\begin{equation}
\Phi(\vec{r},t)=\nabla^{2}\Psi(\vec{r},t),
\end{equation}
and so
\begin{equation}
(E_{0}^2-c^2\hbar^2\nabla^2)\Phi(\vec{r},t)=-(1+\frac{V(x_{\mu})}{E_{0}})^2\hbar^2
\frac{\partial^2}{\partial t^2}\Phi(\vec{r},t),
\end{equation}
The Eq. (66) is  time-dependent relativistic wave equation of the
material particle in the external potential field. Obviously, when
$V(x_{\mu})=0$ the equation becomes the Klein-Gordon equation.

Now, we give an example for a charged particle in electromagnetic
field. Firstly, we give its time-independent relativistic wave
equation and its $\gamma L$ is \cite{s5}
\begin{equation}
\gamma L=-m_{0}c^{2}+eA_{u}U_{u},
\end{equation}
with
$$A_{\mu}=(\vec{A},\frac{i}{c}\varphi), U_{\mu}=\gamma(\vec{v},ic),$$
where $A_{\mu}$ is electromagnetic four-vector, $U_{\mu}$ is
four-velocity.
\\
and so
\begin{equation}
V(x_{u})=-eA_{\mu}U_{\mu}=-e\gamma(\vec{A}\cdot\vec{v}-\varphi),
\end{equation}
also
$$\vec{p}=\gamma m_{0}\vec{v}, \gamma^{2}=\frac{1}{1-\frac{v^{2}}{c^{2}}},$$
then
\begin{equation}
\vec{v}=\frac{c\vec{p}}{\sqrt{p^{2}+m_{0}^{2}c^{2}}}=\frac{c^{2}{\vec{p}}}{\varepsilon},
\end{equation}

\begin{equation}
\gamma=\frac{\sqrt{p^{2}+m_{0}^{2}c^{2}}}{m_{0}c}=\frac{\varepsilon}{E_{0}},
\end{equation}
and so
\begin{equation}
V(x_{\mu})
=-\frac{e}{E_{0}}(c^{2}\vec{A}\cdot\vec{p}-\varepsilon\varphi),
\end{equation}
Making $\vec{A}\cdot\vec{p}$ into operator
\begin{equation}
\vec{A}\cdot\vec{p}\rightarrow\frac{\vec{A}\cdot\hat{\vec{p}}+\hat{\vec{p}}\cdot\vec{A}}{2},
\end{equation}
and
\begin{equation}
\hat{\vec{p}}\cdot\vec{A}-\vec{A}\cdot\hat{\vec{p}}=-i\hbar\nabla\cdot\vec{A},
\end{equation}
so
\begin{equation}
\hat{\vec{p}}\cdot\vec{A}=-i\hbar\nabla\cdot\vec{A}+\vec{A}\cdot\hat{\vec{p}},
\end{equation}
combining the Eq. (72) with Eq. (74) gives
\begin{equation}
\vec{A}\cdot\hat{p}\rightarrow
\vec{A}\cdot\hat{\vec{p}}-\frac{i\hbar}{2}\nabla\cdot\vec{A}
=-i\hbar(\vec{A}\cdot\nabla+\frac{1}{2}\nabla\cdot\vec{A}),
\end{equation}
from the Eq. (71) and Eq. (75)
\begin{equation}
V(x_{\mu})=-\frac{e}{E_{0}}[-i\hbar
c^2(\vec{A}\cdot\nabla+\frac{1}{2}\nabla\cdot\vec{A})-\varepsilon\varphi],
\end{equation}
substitution of the Eq. (76) into Eq. (59) gives
\begin{equation}
-c^2\hbar^2\nabla^2\Psi(\vec{r})=\{[E-\frac{e}{E_0} (i\hbar
c^2(\vec{A}\cdot\nabla+\frac{1}{2}\nabla\cdot\vec{A})+\varepsilon\varphi)]^2-E_0^2\}
\{1+\frac{e}{E_0^2}[i\hbar
c^2(\vec{A}\cdot\nabla+\frac{1}{2}\nabla\cdot\vec{A})+\varepsilon\varphi]\}^2\Psi(\vec{r}),
\end{equation}
The Eq. (77) is  time-independent relativistic wave equation of a
charged particle in electromagnetic field.

In order to get the time-dependent relativistic wave equation of
the material particle in the external potential field, we
substitution of the Eq. (76) into Eq. (66)
\begin{equation}
(E_0^2-c^2\hbar^2\nabla^2)\Psi(\vec{r},t)=-\{1+\frac{e}{E_0^2}
[i\hbar
c^2(\vec{A}\cdot\nabla+\frac{1}{2}\nabla\cdot\vec{A})+\varepsilon\varphi]\}^2\hbar^2
\frac{\partial^2}{\partial t^2}\Psi(\vec{r},t),
\end{equation}
The Eq. (78) is the time-dependent relativistic wave equation of a
charged particle in electromagnetic field.
\\
\\
{\bf 4. the relativistic wave equation of spin $\frac{1}{2}$
particle} \vskip 8pt

In the following, we give the relativistic wave equation of spin
$\frac{1}{2}$ particle in potential field. We know Dirac resolved
the Klein-Gordon wave equation which is second order in space-time
\begin{equation}
\frac{1}{c^2}\frac{\partial^2}{\partial
t^2}\Psi(\vec{r},t)-\nabla^2\Psi(\vec{r},t)+\frac {m_0^2
c^2}{\hbar^2}\Psi(\vec{r},t)=0,
\end{equation}
into first order in space-time
\begin{equation}
\frac{1}{c}\frac{\partial}{\partial
t}\Psi+\vec{\alpha}\cdot\frac{\partial}{\partial
\vec{x}}\Psi+\frac {im_0 c}{\hbar}\beta\Psi=0,
\end{equation}
with
\begin{equation}
\Psi=\left (
\begin {array}{cr}
\Psi_1(r,t)\\
\Psi_2(r,t)\\
\Psi_3(r,t)\\
\Psi_4(r,t)
\end{array}\right)
\end{equation}
The Eq. (80) is the famous Dirac equation. As the same method, we
resolve the time-dependent relativistic wave equation of the
material particle in the external potential field
\\
$$(E_{0}^2-c^2\hbar^2\nabla^2)\Psi(\vec{r},t)=-(1+\frac{V(x_{\mu})}{E_{0}})^2
\hbar^2\frac{\partial^2}{\partial t^2}\Psi(\vec{r},t),$$
\\
i.e.
\begin{equation}
(1+\frac{V(x_{\mu})}{E_{0}})^2\frac{1}{c^2}\frac{\partial^2}{\partial
t^2}\Psi(\vec{r},t)-\nabla^2\Psi(\vec{r},t)+\frac {m_0^2
c^2}{\hbar^2}\Psi(\vec{r},t)=0,
\end{equation}
into
\begin{equation}
(1+\frac{V(x_{\mu})}{E_0})\frac{1}{c}\frac{\partial}{\partial
t}\Psi+\vec{\alpha}\cdot\frac{\partial}{\partial
\vec{x}}\Psi+\frac {im_0c}{\hbar}\beta\Psi=0,
\end{equation}
Premultiply this equation by
\begin{equation}
(1+\frac{V(x_{\mu})}{E_0})\frac{1}{c}\frac{\partial}{\partial
t}\Psi-(\vec{\alpha}\cdot\frac{\partial}{\partial
\vec{x}}\Psi+\frac {im_0c}{\hbar}\beta)\Psi=0,
\end{equation}
Here, the momentum's operator $\frac{\partial}{\partial x_i}$
can't act on the potential function $V(x_(\mu))$, and so
\begin{equation}
(1+\frac{V(x_{\mu})}{E_0})^2\frac{1}{c^2}\frac{\partial^2}{\partial
t^2}\Psi-\sum\limits_{ik}\alpha_i\alpha_k\frac{\partial}{\partial
x_i}\frac{\partial}{\partial x_k}\Psi+\frac
{m_0^2c^2}{\hbar^2}\beta^2\Psi-\frac{imc}{\hbar}
\sum\limits_i(\alpha_i\beta+\beta\alpha_i)\frac{\partial}
{\partial x_i}\Psi=0,
\end{equation}
As the same method by which Dirac obtained his equation, we can
get the same matrix
\begin{equation}
\alpha_i=\left(
\begin {array}{cr}
0 & \sigma_{i}\\
\sigma_{i} & 0
\end {array}  \right),
\beta=\left(
\begin {array}{cr}
I & 0\\
0 & -I
\end {array}  \right),
\end{equation}
from the Eq. (83), we can obtain
\begin{equation}
i\hbar\frac{\partial}{\partial
t}\Psi=\frac{E_0}{E_0+V(x_\mu)}(-i\hbar
c\vec{\alpha}\cdot\frac{\partial}{\partial
\vec{x}}+m_0c^2\beta)\Psi=\hat{H}\Psi,
\end{equation}
with
\begin{equation}
\hat{H}=\frac{E_0}{E_0+V(x_\mu)}(-i\hbar
c\vec{\alpha}\cdot\frac{\partial}{\partial \vec{x}}+m_0c^2\beta)
\end{equation}
The operator $\hat{H}$ is Hamilton operator of spin $\frac{1}{2}$
particle in external field.
\\
for a stationary state
\begin{equation}
\Psi=\Psi(r) e^{-iEt/\hbar}
\end{equation}
The Eq. (87) becomes
\begin{equation}
\frac{E_0}{E_0+V(x_\mu)}(-i\hbar
c\vec{\alpha}\cdot\frac{\partial}{\partial
\vec{x}}+m_0c^2\beta)\Psi=E\Psi
\end{equation}
The Eq. (90) is the eigenvalue equation of energy $E$ of spin
$\frac{1}{2}$ particle in external field.

In the following, we will give the relativistic wave equation of
spin $\frac{1}{2}$ and $m_0=0$ particle in external potential
field. From the Eq. (82), we can get the time-dependent
relativistic wave equation of the material particle in the
external potential field for $m_0=0$
\begin{equation}
(1+\frac{V(x_{\mu})}{E_{0}})^2\frac{1}{c^2}\frac{\partial^2}{\partial
t^2}\Phi-\nabla^2\Phi=0,
\end{equation}
The relativistic wave equation of spin $\frac{1}{2}$ and $m_0=0$
particle is
\begin{equation}
(1+\frac{V(x_{\mu})}{E_0})\frac{1}{c}\frac{\partial}{\partial
t}\Phi+\sum\limits_{i=1}^{3}\sigma_i\frac{\partial}{\partial
x_i}\Phi=0,
\end{equation}
Premultiply this equation by
\begin{equation}
-(1+\frac{V(x_{\mu})}{E_0})\frac{1}{c}\frac{\partial}{\partial
t}\Phi+\sum\limits_{k=1}^{3}\sigma_k\frac{\partial}{\partial
x_k}\Phi=0,
\end{equation}
and so
\begin{equation}
\{-(1+\frac{V(x_{\mu})}{E_0})^2\frac{1}{c^2}\frac{\partial^2}{\partial
t^2}+\sum\limits_{i,k}(\sigma_k\sigma_i\frac{\partial}{\partial
x_k}\frac{\partial}{\partial x_i})\}\Phi=0,
\end{equation}
Symmetrizing the Eq. (94)
\begin{equation}
\{-(1+\frac{V(x_{\mu})}{E_0})^2\frac{1}{c^2}\frac{\partial^2}{\partial
t^2}+\frac{1}{2}\sum\limits_{i,k}(\sigma_k\sigma_i+\sigma_k\sigma_i)\frac{\partial}{\partial
x_k}\frac{\partial}{\partial x_i}\}\Phi=0,
\end{equation}
Comparing the Eq. (95) with Eq. (91), we can get
\begin{equation}
\frac{1}{2}(\sigma_k\sigma_i+\sigma_k\sigma_i)=\delta_{ik},
(i,k=x,y,z)
\end{equation}
so
$$\sigma_x^2=\sigma_y^2=\sigma_z^2=1,$$
$$\sigma_x\sigma_y=-\sigma_y\sigma_x,$$
where $\sigma$ is the Pauli matrices. Then the Eq. (93) can be
written as
\begin{equation}
(1+\frac{V(x_{\mu})}{E_0})\frac{1}{c}\frac{\partial}{\partial
t}\Phi=-\vec{\sigma}\cdot\vec{\nabla}\Phi,
\end{equation}
also
\begin{equation}
i\hbar\frac{\partial}{\partial t}\Phi=\frac{-i\hbar
c\vec{\sigma}\cdot\vec{\nabla}}{(1+\frac{V(x_{\mu})}{E_0})}\Phi
\end{equation}
In our work, we give non-relativistic wave equations which are
different from the Schr\"{o}dinger wave equation, and extend the
Klein-Gordon equation, which includes the external field.
Meanwhile, we obtain the relativistic wave equation of spin
$\frac{1}{2}$ particle in external field. We think our theory
maybe have some effects to quantum field theory.

\end{document}